\documentclass{IOS-Book-Article}

\usepackage{mathptmx}
\usepackage{soul}\setuldepth{article}

\begin{document}

\pagestyle{headings}
\def\thepage{}

\begin{frontmatter}   
\title{Relational Artificial Intelligence}

\author{\fnms{Virginia} \snm{Dignum}%
\thanks{Corresponding Author: Virginia Dignum, Ume{\aa} University, Sweden, E-mail:\url{virginia@cs.umu.se}.\\
This paper is a modified and extended version of my paper `Responsible Artificial Intelligence: recommendations and lessons learned', to appear in the book \textit{`Responsible AI in Africa'}, edited by Damian Okaibedi Eke, Kutoma Wakunuma and Simisola Akintoye.}}

\address{Computing Science Department, Ume{\aa} University, Sweden}

\begin{abstract}
The impact of Artificial Intelligence does not depend only on fundamental research and technological developments, but for a large part on how these systems are introduced into society and used in everyday situations. 
Even though AI is traditionally associated with rational decision making, understanding and shaping the societal impact of AI in all its facets requires a relational perspective. A rational approach to AI, where computational algorithms drive decision making independent of human intervention, insights and emotions, has shown to result in bias and exclusion, laying bare societal vulnerabilities and insecurities. 

A relational approach, that focus on the relational nature of things, is needed to deal with the ethical, legal, societal, cultural, and environmental  implications of AI.
A relational approach to AI recognises that objective and rational reasoning cannot does not always result in the `right’ way to proceed because what is `right’ depends on the dynamics of the situation in which the decision is taken, and that rather than solving ethical \textit{problems} the focus of design and use of AI must be on asking the ethical \textit{question}.

In this position paper, I start with a general discussion of current conceptualisations of AI followed by an overview of existing approaches to governance and responsible development and use of AI. Then, I reflect over what should be the bases of a social paradigm for AI and how this should be embedded in relational, feminist and non-Western philosophies, in particular the Ubuntu philosophy. 
\end{abstract}

\end{frontmatter}   



\section{Introduction}
 
As Artificial Intelligence (AI) is increasingly impacting many aspects of life, the awareness that it has the potential to impact our lives and our world as no other technology has done before is rightfully raising many questions concerning its ethical, legal, societal and economical effects. Moreover, whereas the dangers and risks of application of AI without due consideration of its societal, ethical or legal impact, are increasingly acknowledged, the potential of AI to contribute to human and societal well-being cannot be dismissed. 
But AI means different things to different people.

Ensuring the responsible development and use of AI is becoming a main direction in AI research and practice. Governments, corporations and international organisations alike are coming forward with proposals and declarations of their commitment to an accountable, responsible, transparent approach to AI, where human values and ethical principles are leading. A comprehensive study of the role of AI in achieving the Sustainable Development Goals\footnote{\url{https://sdgs.un.org/}} in which I participated, concluded that it has the potential to shape the delivery of all 17 goals, contributing positively to 134 targets across all the goals, but it may also inhibit 59 targets \cite{vinuesa2020a}. 
Currently, there are over 600 AI-related policy recommendations, guidelines or strategy reports, which have been released by prominent intergovernmental organisations, professional bodies, national-level committees and other public organisations, non-governmental, and private for-profit companies. A recent study of the global landscape of AI ethics guidelines shows that there is a global convergence around five ethical principles: Transparency, Justice and Fairness, Non-Maleficence, Responsibility, and Privacy \cite{jobin2019a}. 

These are much-needed efforts, but still much work is needed to ensure that all AI is developed and used in responsible ways that contribute to trust and well-being. Nevertheless, even though organisations agree on the need to consider ethical, legal and societal principles, how these are interpreted and applied in practice, varies significantly across the different recommendation documents. 

At the same time, the growing hype around `AI’ is blurring its definition and shoving into the same heap concepts and applications of many different sorts. A hard needed first step in the responsible development and use of AI is to ensure a proper AI narrative, one that demystifies its capabilities, minimises both overselling and underselling of AI-driven solutions, and 
that enables wide and inclusive participation in the discussion on the role of AI in society. Understanding the capabilities and addressing the risks of AI, requires that all of us, from developers to policy-makers, from provides to end-users and bystanders, have a clear understanding of what AI is, how it is applied and what are the opportunities and risks involved.


\section{What is AI and why should we care?}
Without a proper understanding of what AI is and what it can, and cannot, do, all the efforts towards governance, regulation and responsible use of AI have the risk to become void. Current AI narratives bring forward benefits and risks and describe AI in many different ways, from the obvious next step in digitisation to some kind of magic. If the `business as usual' narrative is detrimental of innovation and contributes to the maintenance of status quo, the `magic' narrative, well fed from science fiction, Hollywood and the popular press, often supports a feeling that nothing can be done against such an all-knowing entity that rules over us in possibly unexpected ways, either solving all our problems, or destroying the world in the process. In both cases, the danger is that the message is that little can be done against the risks and challenges of AI.

The reality is, as usual, somewhere in the middle. In the following, I briefly describe some of the ways AI is often misunderstood, and conclude with a reflection of the significance to the current efforts towards AI governance.

\subsection{AI is not intelligent }
John McCarthy, who originally coined the term Artificial Intelligence, defined it as ``the study and design of intelligent agents". In this definition, which is still one of the most common definitions of AI, the concept of intelligence refers to the ability of computers to perform tasks commonly associated with intelligent beings, i.e. humans or other non-human animals. 

Currently, AI is mostly associated with Machine Learning (ML) but the field of AI includes many other techniques\footnote{Chapter 2 in my book `Responsible Artificial Intelligence' gives a short, non-technical overview of different techniques \cite{raibook2019}.}. Machine Learning, and in particular, Neural Networks, or Deep Learning, is a subset of AI techniques that uses statistical methods to enable computers to perceive some characteristics of their environment. Current techniques are particularly efficient in perceiving images, written or spoken text, as well as the many applications of structured data. By analysing many thousands of examples (typically a few million), the system is able to identify commonalities in these examples, which then enable it to interpret data that it has never seen before, which is often referred to as prediction. 

The question remains of what is human (or animal) intelligence, and whether intelligence can be achieved by correlation techniques \cite{Larson2021}. However, there is no single accepted definition of intelligence, even though in cognitive science, intelligence is associated with the ability of the mind to reach correct conclusions about what is true and what is false, and about how to solve problems \cite{colman2015a}. As best, we can say that intelligence is a multifaceted concept. Psychologists debate on issues such as types of intelligence, the role of nature versus nurture in intelligence, how intelligence is represented in the brain, and the meaning of group differences in intelligence. Major theories include Sternberg’s triarchic theory, Gardner’s theory of multiple intelligences \cite{gardner2011a}, and Piaget’s theory of development \cite{piaget1964a}. Many characterise human intelligence as more than an analytical process and to include creative, practical and other abilities. These abilities, for a large part associated with socio-cultural background and context, are far from being possible to be replicated by AI systems, even if these may approach analytical intelligence for some (simple) tasks. 

\subsection{AI is not magic}
AI is not magic and it is not only artificial. By itself, it will not solve all our problems, nor can it exist without the use of natural resources and the work and efforts of legions of people. In a recent book, ‘The Atlas of AI’ \cite{crawford2021a}, Kate Crawford describes the field as a collection of maps that enable the reader to traverse places, their relations and their impact on AI as an infrastructure. From the mines where the core components of hardware originate, to the warehouses where human labourers are mere servants to the automated structure, in an uneasy reminder of Chaplin’s Modern Times, to the hardship of data classification by low paid workers in data labelling farms, Crawford exposes the hard reality of the hidden side of AI success, positioning that AI is not only not intelligent, it is also not artificial. Concluding with the powerful reminder that AI is not an objective, neutral and universal computational technique, Crawford reminds us that AI is deeply embedded in the social, political, cultural and economic reality of those that build, use and mostly control it \cite{dignum2021a}. 

\subsection{AI is not the algorithm}
AI is based on algorithms, but then so is any computer program and most of the technologies around us. Nevertheless, the concept of `algorithm’ is achieving magical proportions, used right and left to signify many things, \textit{de facto} seen as a synonym to AI. 

The easiest way to understand an algorithm is as a recipe, a set of precise rules to achieve a certain result. Every time you multiply two numbers, you are using an algorithm, as well as you are when you are baking an apple pie. However, by itself, the recipe has never turned into an apple pie; and, the end result of your pie has as much to do with your baking skills and your choice of ingredients, as with the choice for a specific recipe. The same applies to AI algorithms: for a large part the behaviour and results of the system depends on its input data, and on the choices made by those that developed, trained and selected the algorithm. In the same way as we have the choice to use organic apples to make our pie, in AI we also have the choice to use data that respects and ensures fairness, privacy, transparency and all other values we hold dear. This is what Responsible AI is about, and includes demanding the same requirements from the ones that develop the systems that affect us.  

\subsection{AI is a socio-technical system}
If AI is not intelligent, nor magic, nor business as usual, nor an algorithm, how best can we describe AI in order to take into account not only its capabilities but also its societal implications?

AI is first and foremost technology that can automatise (simple, lesser) tasks and decision making processes. At the present, AI systems are largely incapable of understanding meaning. An AI system can correctly identify cats in videos or cancer cells in scan images, but it has no idea of what a cat or a cancer cell is. Moreover, AI system can only do this if there are enough people performing the tasks (classification, collection, maintenance...) that are needed to make the system function, misleadingly, in an autonomous manner. 

At the same time, considering its societal impact and need for human contribution, AI is much more than an automation technique. In this sense, AI can best be understood as a socio-technical ecosystem, recognising the interaction between people and technology, and how complex infrastructures affect and are affected by society and by human behaviour. 
As such, AI is not just about the automation of decisions and actions, the adaptability to learn from the changes affected in the environment, and the interactivity required to be sensitive to the actions and aims of other agents in that environment, and decide when to cooperate or to compete. It is mostly about the structures of power, participation and access to technology that determine who can influence which decisions or actions are being automated, which data, knowledge and resources are used to learn from, and how interactions between those that decide and those that are impacted are defined and maintained. 

A responsible, ethical, approach to AI will ensure transparency about how adaptation is done, responsibility for the level of automation on which the system is able to reason, and accountability for the results and the principles that guide its interactions with others, most importantly with people. In addition, and above all, a responsible approach to AI makes clear that AI systems are artefacts manufactured by people for some purpose, and that those which make these have the power to decide on the use of AI. It is time to discuss how power structures determine AI and how AI establishes and maintains power structures, and on the balance between, those who benefit from, and those who are harmed by the use of AI \cite{crawford2021a}. 

\section{Ensuring the responsible development and use of AI }
Ethical AI is not, as some may claim, a way to give machines some kind of `responsibility’ for their actions and decisions, and in the process, discharge people and organisations of their responsibility. On the contrary, ethical AI gives the people and organisations involved more responsibility and more accountability: for the decisions and actions of the AI applications, and for their own decision of using AI in a given application context. When considering effects and the governance thereof, the technology, or the artefact that embeds that technology, cannot be separated from the socio-technical ecosystem of which it is a component. Guidelines, principles and strategies to ensure trust and responsibility in AI, must be directed towards the socio-technical ecosystem in which AI is developed and used. It is not the AI artefact or application that needs to be ethical, trustworthy, or responsible. Rather, it is the social component of this ecosystem that can and should take responsibility and act in consideration of an ethical framework such that the overall system can be trusted by the society. Having said this, governance can be achieved by several means, softer or harder. Currently several directions are being explored, the main ones are highlighted in the remainder of this section. Future research and experience will identify which approaches are the most suitable, but given the complexity of the problem, it is very likely that a combination of approaches will be needed. 

\subsection{Regulation}
AI regulation is a hot topic, with many supporters and opponents. The recent AI Act proposed by the European Commission envisions a risk-based approach to regulation that ensures that people can trust that AI technology is used in a way that is safe and compliant with the law, including the respect of fundamental human rights\footnote{See \url{https://eur-lex.europa.eu/legal-content/EN/TXT/?uri=CELEX\%3A52021PC0206}.}. The draft proposes to implement most of the seven requirements put forward in the Ethics Guidelines for Trustworthy AI as specific requirements for `high-risk’ AI applications. However, it does not deal explicitly with issues of inclusion, non-discrimination and fairness. 
Minimising or eliminating discriminatory bias or unfair outcomes is more than excluding the use of low-quality data. The design of any artefact, such as an AI system, is in itself an accumulation of choices and choices are biased by nature as they involve selecting an option over another. Technical solutions at dataset level must be complemented by socio-technical processes that help avoid any discriminatory or unfair outcomes of AI. 

Successful regulation demands clear choices about what is being regulated: the AI Act draft does not yet clearly indicate if it is mostly focusing on AI technology, or on the impact and results of its application? By defining AI systems as using `machine learning, logic, or statistical approaches’ it opens the door to endless semantic debates. Not only is the line between old-fashioned
computing and AI irrelevant, it is also blurry. What matters is whether it poses risks to people, society and/or environment. 

A future-proof regulation should focus on the outcomes of systems, whether or not these systems fall in the current understanding of what is ‘AI’. If someone is wrongly identified, is denied human rights or access to resources, or is conditioned to believe or act in a certain way, it does not matter whether the system is ‘AI’ or not. It is simply wrong. Moreover, regulation must also address the inputs, processes and conditions under which AI is developed and used are at least as important. Much has been said about the dangers of biased data, and discriminating applications. Attention for the societal, environmental and climate costs of AI systems is increasing. All these must be included in any effort to ensure the responsible development and use of AI. 

At the same time, AI systems are computer applications, i.e. are artefacts, and as such subject to existing constraints, legislation, for which due diligence obligations and liabilities apply. That is, already now, AI does not operate in a lawless space. Before defining extra regulations, we need to start by understanding what is already covered by existing legislation. 

A risk-based approach to regulation, as proposed by the European Commission, is the right direction to take, but needs to be informed by a clear understanding of what is the source of those risks. Moreover, it requires to not merely focus on technical solutions at the level of the algorithms or the datasets, but rather on developing socio-technical processes, and the corporate responsibility, to ensure that any discriminatory or unfair outcomes are avoided and mitigated. Independently of whether we call the system ‘AI’ or not. 

\subsection{Standardisation }
Standards are consensus-based agreed-upon ways of doing things by providing what they consider to be the minimum universally-acknowledged specifications. Industry standards are proven to be beneficial to organisations and individuals. Standards can help reduce costs and improve efficiency of organisations by providing consistency and quality metrics, the establishment of a common vocabulary, good-design methodologies, and architectural frameworks. At the same time, standards provide consumers with confidence in the quality and safety of products and services. 
Most standards are considered soft governance; i.e. non-mandatory to follow. Yet, it is often in the best interest of companies to follow them to demonstrate due diligence and, therefore, limit their legal liability in case of an incident. Moreover, standards can ensure user-friendly integration between products \cite{theodorou2020a}. 
AI standards work to support the governance of AI development and use is ongoing at ISO and IEEE, the two leading standards bodies. Such standards can support AI policy goals in particular where it concerns safety, security and robustness of AI, guarantees of explainability, and means to reduce bias in algorithmic decisions \cite{cihon2019a}. 
Jointly with the International Electrotechnical Commission (IEC), ISO has established a Standards Committee on Artificial Intelligence (SC-42)\footnote{See \url{https://www.iso.org/committee/6794475.html}.} but ongoing SC-42 efforts are, so far, limited and preliminary \cite{cihon2019a}. 

At the same time, IEEE’s Standards Association global initiative on Ethically Aligned Design\footnote{See \url{https://ethicsinaction.ieee.org/}.} is actively working on vision and recommendations to address the values and intentions as well as legal and technical implementations of autonomous and intelligent systems to prioritise human well-being \cite{chatila2019ieee}. This is the joint work of over 700 international researchers and practitioners. In particular, the P7000  series aims to develop standards that will eventually serve to underpin and scaffold future norms and standards within a new framework of ethical governance for AI/AS design. Currently, the P7000 working groups are working on candidate standard recommendations to address issues as diverse as system design, transparency in autonomous systems, algorithmic bias, personal, children, student, and employer data governance, nudging, or, the identification and rating the trustworthiness of news sources. Notably, the efforts on assessment of impact of autonomous and intelligent systems on human well-being is now available as an IEEE standard \cite{p7000}. 

\subsection{Assessment}
Responsible AI is more than the ticking of some ethical `boxes’ or the development of some add-on features in AI systems. Nevertheless, developers and users can benefit from support and concrete steps to understand the relevant legal and ethical standards and considerations when making decisions on the use of AI applications. Impact assessment tools provide a step-by-step evaluation of the impact of systems, methods or tools on aspects such as privacy, transparency, explanation, bias, or liability \cite{taddeo2018a}. 
It is important to realise that even though these approaches \textit{``can never map the entire spectrum of opportunities, risks, and unintended consequences of AI systems, they may identify preferable alternatives, valuable courses of action, likely risks, and mitigating strategies. This has a dual advantage. As an opportunity strategy, foresight methodologies can help leverage ethical solutions. As a form of risk management, they can help prevent or mitigate costly mistakes, by avoiding decisions or actions that are ethically unacceptable”} \cite{taddeo2018a}.

Currently, much effort is being put on the development of assessment tools. The EU Guidelines for trustworthy AI are accompanied by a comprehensive assessment framework which was developed based on a public consultation process. 
Finally, it is important to realise that any requirements for trustworthy AI are necessary but not sufficient to develop human-centred AI. That is, such requirements need to be understood and implemented from a contextual perspective, i,e, it should be possible to adjust the implementation of the requirement such as transparency based on the context in which the system is used. I.e. requirements such as transparency should not have one fixed definition for all AI systems, but rather be defined based on how the AI system is used. At the same time, any AI technique used in the design and implementation should be amenable to explicitly consider all ethical requirements. E.g. it should be possible to explain (or to show) how the system got to a certain decision or behaviour. 

Assessment tools need to be able to account for this contextualisation, as well as ensuring alignment with existing frameworks and requirements in terms of other types of assessment, such that the evaluation of trust and responsibility of AI systems provides added value to those developing and using it, rather than adding yet another bureaucratic burden. 

\subsection{Codes of conduct and advisory boards}
A professional code of conduct is a public statement developed for and by a professional group to reflect shared principles about practice, conduct and ethics of those exercising the profession; describe the quality of behaviour that reflects the expectations of the profession and the community; provide a clear statement to the society about these expectations, and enable professionals to reflect on their own ethical decisions. 
A code of conduct supports professionals to assess and resolve difficult professional and ethical dilemmas. While there in the case of ethical dilemmas there is not a correct solution, the professionals can give account of their actions by referring to the code. In line with other socially sensitive professions, such as medical doctors or lawyers, i.e., with the attendant certification of ‘ethical AI’ can support trust. Several organisations are working on the development of codes of conduct for data and AI related professions, with specific ethical duties. Just recently ACM, the Association for Computing Machinery, the largest international association of computing professionals, updated their code of conduct. This voluntary code is `\textit{`a collection of principles and guidelines designed to help computing professionals make ethically responsible decisions in professional practice. It translates broad ethical principles into concrete statements about professional conduct\footnote{\url{https://www.acm.org/code-of-ethics}}”}. This code explicitly addresses issues associated with the development of AI systems’, namely issues of emergent properties, discrimination and privacy. Specifically, it calls out the responsibility of technologists to ensure that systems are inclusive and accessible to all, and requires that they are knowledgeable about privacy issues. 

Many organisations have since established the role of chief AI ethics officer, or similar. Others, recognising that the societal and ethical issues that arise from AI are complex and multi-dimensional, and therefore require insights and expertise from many different disciplines and an open participation of different stakeholders, have established AI ethics boards or advisory panels. These roles and organisational components, are becoming a hot topic as large businesses are increasingly dependent on AI and as the impact of these systems on people and society becomes increasingly more evident, and not always for the best. Recent scandals both about the impact of AI in bias and discrimination, as on the way businesses are dealing with their own responsibility, specifically on the role and treatment of whistle-blowers, have increased the demand for clear and explicit organisational structures to deal with the impact of AI. 

\subsection{Awareness and Participation }
Inclusion and diversity are a broader societal challenge and central to AI development. It is therefore important that as broad a group of people as possible have a basic knowledge of AI, what can (and can’t) be done with AI, and how AI impacts individual decisions and shapes society. A well-known initiative in this area is Elements of AI\footnote{See \url{https://www.elementsofai.com/}.}, initiated in Finland with the objective to train one percent of EU citizens in the basics of artificial intelligence, thereby strengthening digital leadership within the EU. 

In parallel, research and development of AI systems must be informed by diversity, in all the meanings of diversity, and obviously including gender, cultural background, and ethnicity. Moreover, AI is not any longer an engineering discipline and at the same time there is growing evidence that cognitive diversity contributes to better decision making. Therefore, developing teams should include social scientists, philosophers, and others, as well as ensuring gender, ethnicity and cultural differences. It is equally important to diversify the discipline background and expertise of those working on AI to include AI professionals with knowledge of, amongst others, philosophy, social science, law and economy. Regulation and codes of conduct can specify targets and goals, along with incentives, as a way to foster diversity in AI teams \cite{dignum2020a}. 

\section{From a rational to a relational conception of AI}
 
The dominant approach to AI has so far been an individualistic, rational one. Russell and Norvig’s classic AI text book, defines AI along two dimensions \cite{russell2010a}: how it reasons (human-like or rationally ) and what it `does’ (think or act). This classification does not mean that human reasoning is not rational, but in their definition, human-like approaches aim to understand and model how the human mind works, and rational approaches aim developing systems that that result in the optimal level of benefit or utility for an individual. Both approaches are well aligned with the central element of Western philosophy ``Cogito ergo sum" (I think therefore I am), and fundamentally conceptualise an AI system as an individual entity capable of reasoning.

Throughout the whole history of AI, intelligent agents are  characterised as bounded rational, acting towards their own perceived interests \cite{simon1991bounded}. For instance, by identifying and applying patterns in (human-generated) data, machine learning systems mimic and extend the human reasoning and actions embedded in that data, whereas symbolic logic approaches (the so called ‘good old-fashioned AI’, or GOFAI) aim to capture the laws of rational thought and action, resulting in an idealised model of human reasoning. In both cases, the view is that of an entity that reasons...

Rationality is often a central assumption for agent deliberation \cite{russell2010a,dignum2017a}. Moreover, intelligent systems are expected to hold consistent world views (beliefs), and to optimise action and decision based on a set of given preferences (often accuracy has highest priority). This view on rationality entails that agents are expected, and designed, to act rationally in the sense that they choose the best means available to achieve a given end, and maintain consistency between what is wanted and what is chosen \cite{lindenberg2001a}. 

The main advantages of a rationality assumption are their parsimony and applicability to a very broad range of situations and environments, and their ability to generate falsifiable, and sometimes empirically confirmed, hypotheses about actions in these environments. This gives conventional rational choice approaches a combination of generality and predictive power not found in other approaches. 

Both the human-like and the rational perspectives on AI described in \cite{russell2010a} are suited for a task-oriented view on the purpose of AI systems. That is, for situations where the system is expected to optimise the result of its actions for a specific purpose. 
Unfortunately, this type of  reasoning mostly fits those situations 
where all required information is available or can be gathered at will. It does not really suit most human behaviour which is based on split-second decisions, on habits, on social conventions and on power structures. Moreover, even when the AI agent is able to perceive its environment and adapt accordingly, it is mostly unaware of its own role in that environment, and of the fact that its actions contribute to change, and therefore unable to handle the result of its action.

Given the large impact of AI on society, a new modelling paradigm is needed that is able to account for a feedback loop of decision--action that is fundamentally grounded on context. When the aim of AI systems is to align with societal behaviour and to develop systems that are able to interact with people in social settings, such rationality view is not enough to model human behaviour. That is, AI modelling needs to follow a social paradigm 
that can account for the reality in which human behaviour is neither simple nor rational, but derives from a complex mix of mental, physical, emotional and social aspects. Realistic AI applications must therefore consider situations in which not all alternatives, consequences, and event probabilities can be foreseen. It is impossible to ``rationally" optimise utility, as the utility function is not completely known, neither are the optimisation criteria known. This renders rational choice approaches unable to accurately model and predict a wide range of human behaviours. Already in 2010, we described how different types of variations and models cater for different applications, while no generic model exists that serves as a foundation for all models \cite{dignum2010a}.

\section{Elements of a relational paradigm for AI}
AI, like all technology, affects and changes our world, which in turn changes us. New paradigms are needed that address collective understanding and the effect of change in context and the feedback loop from change back to the individual and collective reasoning and behaviour. Modelling this feedback loop recognises that it is not just about which action is performed but, what kind of reasoning leads to that action, and which values and perceptions lead the observation of the context. Current AI paradigms 
strengthen power structures and prevent a truly societal understanding of the impact and challenges of AI for humanity and society. Theories and models to support the understanding of how AI is \textit{experienced} by individuals and social groups, rather than how it is designed or operationalised, are mostly lacking.

Given the impact AI systems can have on people, inter-personal interactions, and society as a whole, it seems to be relevant to consider a relational stance to approach the specification, development and analysis of AI systems. Both feminist theories and non-Western philosophies (in particular the Ubuntu philosophy) take a societal rather than an individual stance, which may provide a suitable grounds for a relational conceptualisation of AI. 
In this section, I will reflect on possible ways on how Ubuntu and feminist philosophies could serve this purpose. 

\subsection{Ubuntu philosophy and AI}
Ubuntu expresses the deeply-held African ideals of one’s identity being rooted in one’s interconnectedness with others, and emphasises norms for interpersonal relationships that contribute to social justice, such as reciprocity, selflessness and symbiosis. Community is at the core of Ubuntu, focusing on interconnectedness and caring for communal living, underpinned by values of cooperation and collaboration \cite{mugumbate2013a}. Solidarity, which requires people to be aware of and attentive to the needs of those around them, rather than focusing only on their own needs is therefore central in Ubuntu, with an emphasis on caring, care-taking, and context \cite{breda2019a}. 
As such, Ubuntu philosophy is essentially relational, and defines morally right actions as those that that connect, rather than separate (i.e. honours communal relationships, reduces discord or promotes friendly relationships. The concept of community can best be understood as an (objective) standard that should guide what the majority wants, or what moral norms become central \cite{ewuoso2019a}. This does not imply that individual rights are subordinated but that individuals pursue their own good through pursuing the common good \cite{lutz2009a}. 

Human rights set the foundation of human dignity in terms of autonomy. This view, for a large part originating from Kantian philosophy, sets human rights as the ultimately ways of treating our intrinsically valuable capacity for self-governance with respect. It has therefore been argued that the collectivist grounds of Ubuntu thought are at odds with this individual autonomy view. According to Metz, \textit{``While the Kantian theory is the view that persons have a superlative worth because they have the capacity for autonomy, the present, Ubuntu-inspired account is that they do because they have the capacity to relate to others in a communal way."} \cite[p.544]{metz2011a}. Or, as Metz also describes: \textit{``Human rights violations are ways of gravely disrespecting people’s capacity for communal relationship, conceived as identity and solidarity [...]"} \cite[p.545]{metz2011a}. In Ubuntu, human nature is special and inviolable due to its capacity for harmonious relationships. At the same time, no individual’s rights are greater than another, thus, every individual in a community, including both children and adults is important and should be heard and respected \cite{osei-hwedie2007a}. 

With respect to the ethics of AI development and use, the above formulation of human dignity as the human capability to relate to others in a communal way, can account for, or justify, the resolution of moral dilemmas, where autonomy conflicts with beneficence or any of the other principles, as also proposed with respect to bioethics and medicine, or to ground the UN’s sustainable development goals\cite{ewuoso2019a}. 

Specifically, the formulation of Ubuntu described above may be used to justify decisions in face of ethical dilemma(s), for example where such a decision favours the action that enhances communal relationships, or the capacity for the same \cite{ewuoso2019a}. As such, this framework could usefully supplement utilitarian, individualistic and deontological approaches that are often embedded in AI ethics decision-making. As proposed for the case of clinical contexts, ethical decision making in the context of AI systems, can also be extended with rules that state that ``[a] breach of an ethical principle is justifiable if, on the balance of probabilities, such a breach is more likely to enhance communal relationships (...)" \cite{ewuoso2019a}. 

Given the transnational character of AI, it is imperative to address the ways in which AI may impact or be accepted by society in various regions around the world. In particular, it is needed to position the African continent in global debates and policy-making in responsible AI. For instance, initiatives such as Responsible AI Network – Africa , and the African Observatory on Responsible AI  are aiming at understand how AI may impact or be accepted by society in various regions around the world, deepen the understanding of AI and its effects in (Sub-Saharan) Africa, and promoting the development and implementation of locally appropriate evidence-led AI policies and enabling legislation. 

It is as important to extend current conceptualisations of AI, with the relational world view that characterise African thought, such as embedded in Ubuntu philosophy. Current AI paradigms strengthen existing power structures and prevent a truly societal understanding of the impact and challenges of AI for humanity and society. Given the impact AI systems can have on people, inter-personal interactions, and society as a whole, it seems to be relevant to consider a relational stance to approach the specification, development and analysis of AI systems. The deeply-held African vision of one’s identity being rooted in connectedness to others and to society, as expressed in Ubuntu philosophy, can support integrating such a social perspective to AI – in terms of how it reasons (human-like or rationally) and what it ‘does’ (think or act), which would result in a new paradigm that considers social/collective reasoning and includes change, or reaction, as a third possible result of AI, next to thinking and acting. Figure 1 depicts this perspective on AI, extending the well-known dimensions defined in \cite{russell2010a}, here depicted in grey shading. 

\subsection{Feminist theory and AI}

As discussed in Section 2, the development, use and impact of AI is neither `artificial’ nor `intelligent’ but the product of choices involving theory and values, strongly related by power structures that support and enforce an increasing divide between the `haves' and the `have-nots'. 
Current AI paradigms, as seen in Section 4, rely, and are bound, by individualistic theories of intelligence, thinking, rationality, and human nature but lack the means to deal with values, social norms and collective reasoning, and how these ground and influence human thinking and action. In social action, outcomes are relative to the actions of others, where others are not just seen as opponents or obstacles on the decision making, but as a positive force for achieving a joint endeavour.

Already in 1995, Alison Adam proposed feminist epistemology as a radically different, and qualitative, alternative to traditional views on the AI narrative, centred on the question of whether we are prepared to accept and use such systems in our lives, rather than reflecting on what constitutes intelligence. A feminist approach analyses how AI systems may or may not reinforce existing power structures, and how AI can or cannot represent differing social groups \cite{adam1995artificial}. More recently, Catherine D'Ignazio stressed this need to  think critically about the politics and ethics of representation \cite{dignazio2015would}. Drawing a parallel with the role of cartographic maps as instruments of power, producing worlds that are intimately bound up with that power, data analysis and visualisation are also socially situated knowledge, ultimately determining our interaction with and understanding of the AI system. 

Many current approaches to ethics in AI approach decision in black and white, good versus bad, under the idea that if only we can model ethics in some formal, digital way, the problem would be solved,  the AI system `be' ethical, and (other) people will be wiser and better human beings. What is usually known as `Ethics by design' has a very techno-solutionist ring to it... Reality is messier; common-sense morality, moral indignation and assignment of guilt and blame damage relationships, create separateness, and narrow our scope of possible actions. Classic ethics theories are too abstract, practical application leads to too many exceptions, and each can lead to a very different `solution' to any given dilemma. 

Research on algorithmic injustice shows how AI and in particular Machine Learning automates and perpetuates historical, often unjust and discriminatory, patterns. The negative consequences of algorithmic systems, especially on marginalized communities, require a deeper understanding of the context in which AI systems are designed, developed and used, that needs go much further than debiasing and/or making datasets inclusive \cite{birhane2021algorithmic}.
Relational ethics is borne out of a feminist ethics of care and recognises that objective and rational moral reasoning cannot always advise on the `right’ way to proceed – what is `right’ emerges within the dynamics of research relationships and can depend on the time, place and people involved. Relational ethics moves away from attempting to solve the ethical `problem’ to asking the ethical `question’.


At the same time, a relational stance to AI also addresses collectivist views and the continuous loop between change and decision. Modelling this feedback loop recognises that the development and use of AI is not just about the optimisation of performance but, most importantly, the societal and individual context of the values and perceptions and reasoning that lead to action. Addressing change from the human-like perspective, leads to approaches to AI that aim at enhance, rather than replace, human performance, and from a rational perspective, it concerns the development and optimisation of institutional infrastructures that maximise the effects of rational behaviour. 

All these perspectives need to be brought together to address the impact of AI for a socially grounded and engaged perspective. This is no easy feat, but one for which there are no single models, nor simple approaches. It will require multidisciplinary and multi-stakeholder participation and to accept that any solution is always contingent and contextual.



\section{Conclusions}
Increasingly, AI systems will be taking decisions that affect our lives, in smaller or larger ways. In all areas of application, AI must be able to take into account societal values, moral and ethical considerations, weigh the respective priorities of values held by different stakeholders and in multicultural contexts, explain its reasoning, and guarantee transparency. As the capabilities for autonomous decision-making grow, perhaps the most important issue to consider is the need to rethink responsibility. Being fundamentally tools, AI systems are fully under the control and responsibility of their owners or users. However, their potential autonomy and capability to learn, require that design considers accountability, responsibility and transparency principles in an explicit and systematic manner. The development of AI algorithms has so far been led by the goal of improving performance, leading to opaque black boxes. Putting human values at the core of AI systems calls for a mind-shift of researchers and developers towards the goal of improving transparency rather than performance, which will lead to novel and exciting techniques and applications. In particular, this requires to complement the currently predominant individualistic view of AI systems, to one that acknowledges and incorporates the collective, societal, 

Biological evolution has long been revised from a `ladder’ view: a uni-linear progression from ‘primitive’ to ‘advanced’. The same revision is also seen in anthropology: the idea that cultural evolution follows a ladder model, with small-scale decentralised societies at the bottom and hierarchical, state, societies at the top, where the top would be technologically more advanced, has been shown to be not only demeaning but also inaccurate \cite{eglash1999a}. These fields have long since moved to a more dynamic, branching type model that account to interrelations between cooperation and competition, individualism and collectivism, power and influence, cause and effect. Still, the main perspective in AI, is that intelligence is linear, and by creating increasingly more intelligent systems, we will solve the problems of previous ones. It is high time that AI models embrace such a branching view. Only then, can AI align with the diversity that truly reflects worldwide differences in cultural and philosophical thought. 

Rethinking AI from a relational, feminist, non-Western perspective is not a fad or a thought experiment for philosophers. It is ultimately the only way forward, for AI as a scientific field, and more importantly for all of us, and for the world.

\subsubsection*{Acknowledgements}
This work was partially supported by the Wallenberg AI, Autonomous Systems and Software Program (WASP), funded by the Knut and Alice Wallenberg Foundation and by the European Commission’s Horizon 2020 Research and Innovation Programme project HumaneAI-Net (grant 825619).

\bibliographystyle{plain}
\bibliography{refs}

\end{document}